\documentclass[%
 aip,
 amsmath,amssymb,
 reprint,%
]{revtex4-1}

\usepackage{graphicx}% Include figure files
\usepackage{dcolumn}% Align table columns on decimal point
\usepackage{bm}% bold math
\usepackage[utf8]{inputenc}
\usepackage[T1]{fontenc}
\usepackage{mathptmx}
\usepackage{etoolbox}
\usepackage{natbib}
\usepackage{xcolor}

\makeatletter
\def\@email#1#2{%
 \endgroup
 \patchcmd{\titleblock@produce}
  {\frontmatter@RRAPformat}
  {\frontmatter@RRAPformat{\produce@RRAP{*#1\href{mailto:#2}{#2}}}\frontmatter@RRAPformat}
  {}{}
}%
\makeatother
\begin{document}

\preprint{AIP/123-QED}

\title{Enhanced optical properties of MoSe$_2$ grown by molecular beam epitaxy on hexagonal boron nitride}
% Force line breaks with \\
\author{C. Vergnaud}
 \affiliation{Univ. Grenoble Alpes, CEA, CNRS, Grenoble INP, IRIG-Spintec, 38000 Grenoble, France}%Lines break automatically or can be forced with \\
\author{V. Tiwari}%
% \email{Second.Author@institution.edu.}
\affiliation{Univ. Toulouse, INSA-CNRS-UPS, LPCNO, 31077 Toulouse, France
}%
\author{L. Ren}%
\affiliation{Univ. Toulouse, INSA-CNRS-UPS, LPCNO, 31077 Toulouse, France
}%
\author{T. Taniguchi}%
\affiliation{National Institute for Materials Science,
Tsukuba 305-0047, Ibaraki, Japan
}%
\author{K. Watanabe}%
\affiliation{National Institute for Materials Science,
Tsukuba 305-0047, Ibaraki, Japan
}%
\author{H. Okuno}
\affiliation{Univ. Grenoble Alpes, CEA, IRIG-MEM, 38000 Grenoble, France
}%
\author{I. Gomes de Moraes}
\affiliation{Univ. Grenoble Alpes, CEA, CNRS, Grenoble INP, IRIG-Spintec, 38000 Grenoble, France}%
\author{A. Marty}
\affiliation{Univ. Grenoble Alpes, CEA, CNRS, Grenoble INP, IRIG-Spintec, 38000 Grenoble, France}%
\author{C. Robert}%
\affiliation{Univ. Toulouse, INSA-CNRS-UPS, LPCNO, 31077 Toulouse, France
}%
\author{X. Marie}%
\affiliation{Univ. Toulouse, INSA-CNRS-UPS, LPCNO, 31077 Toulouse, France
}%
\author{M. Jamet}
 \affiliation{Univ. Grenoble Alpes, CEA, CNRS, Grenoble INP, IRIG-Spintec, 38000 Grenoble, France}%
 \email{matthieu.jamet@cea.fr.}
 %\homepage{http://www.Second.institution.edu/~Charlie.Author.}
%\affiliation{%
%Second institution and/or address%\\This line break forced% with \\
%}%

\date{\today}% It is always \today, today,

\begin{abstract}
Transition metal dichalcogenides (TMD) like MoSe$_2$ exhibit remarkable optical properties such as intense photoluminescence (PL) in the monolayer form. 
%PL represents a ideal tool to probe proximity effects optically when the TMD monolayer is inserted into a van der Waals heterostructure. However, such study requires narrow-linewidth PL to finely detect proximity effects.
To date, narrow-linewidth PL is only achieved in micrometer-sized exfoliated TMD flakes encapsulated in hexagonal boron nitride (hBN). In this work, we develop a growth strategy to prepare monolayer MoSe$_2$ on hBN flakes by molecular beam epitaxy in the van der Waals regime. It constitutes the first step towards the development of large area single crystalline TMDs encapsulated in hBN for potential integration in electronic or opto-electronic devices. For this purpose, we define a two-step growth strategy to achieve monolayer-thick MoSe$_2$ grains on hBN flakes. The high quality of MoSe$_2$ allows us to detect very narrow PL linewidth down to 5.5 meV at 13 K, comparable to the one of encapsulated exfoliated MoSe$_2$ flakes. Moreover, sizeable PL can be detected at room temperature as well as clear reflectivity signatures of A, B and charged excitons. 
\end{abstract}

\maketitle

%\section{\label{sec:level1}First-level heading:\protect\\ The line
%break was forced \lowercase{via} \textbackslash\textbackslash}

%\section{Introduction}

Hexagonal boron nitride (hBN) has been identified as a key material to encapsulate 2D materials and enhance their electronic properties such as carrier mobility \cite{Mayorov2011} or photoluminescence (PL) \cite{Cadiz2017}. Moreover, for its flatness, inertness and large bandgap, hBN constitutes an ideal substrate for the van der Waals epitaxy of other 2D materials like transition metal dichalcogenides (TMD) of general formula MX$_2$ with M=Mo, W and X=S, Se. It also exhibits a 3-fold symmetry matching the one of TMDs in the 2H phase, which could prevent the formation of twin domains and boundaries to reach high quality TMD monolayers. Previous works reported the molecular beam epitaxy (MBE) growth of MoSe$_2$ monolayers on hBN flakes exfoliated from the bulk crystal \cite{Poh2018,Pacuski2020}. In Ref.\citenum{Poh2018}, Poh et al. have grown MoSe$_2$ monolayer on hBN at two different temperatures 250°C and 500°C, at deposition rates of 0.6 and 1.3 ML/h respectively while keeping a constant Se:Mo ratio of 20. They found that the film grown at 500°C exhibit highly oriented large grains that coalesced to form continuous film. They could measure the photoluminescence at room temperature showing the neutral exciton line with a full width at half maximum (FWHM) of $\approx$60 meV at room temperature. In Ref.\citenum{Pacuski2020}, Pacuski et al. have shown very narrow photoluminescence lines of MBE grown MoSe$_2$ monolayer (6.6 meV for the neutral exciton line at 10 K) demonstrating the ability of MBE to synthesize high quality TMD monolayers with electronic properties comparable to or even better than that of exfoliated TMD flakes. In this work, the authors proceeded in multi-steps alternating growths at 300°C to favor nucleation and annealings at 750°C to improve the crystal quality and clean the surface from the second layer. The Se:Mo ratio was between 100 and 1000.\\
Here, we adopted another growth strategy. The full growth of one MoSe$_2$ layer on exfoliated hBN flakes was achieved at high temperature (800°C) in order to target both high crystalline quality and a limited formation of bilayers. However, due to the ultralow surface energy of hBN, the nucleation rate is negligible at such high temperature. To circumvent this issue, we optimized the nucleation step at 300°C with a low deposition rate. By this, we obtained a reasonable density of monolayer-thick nuclei at the hBN surface and we could proceed with the growth of the MoSe$_2$ monolayer at 800°C at very low deposition rate. We obtained neutral exciton photoluminescence linewidths of the order of 5.5 meV at 13 K, comparable to the one obtained in Ref.\citenum{Pacuski2020} as well as clear signature of A, B and charged excitons in the reflectivity spectrum. Finally, we detected a sizeable photoluminescence signal at room temperature comparable to the one of mechanically exfoliated flakes with a FWHM of the order of 40 meV \cite{Cadiz2017}. This work demonstrates the potential of MBE to grow high quality TMD monolayers on large areas constituting model systems to further study optically proximity effects.\\

%\section{Methods}

The sample preparation starts with the mechanical exfoliation of hBN flakes from high quality crystals (NIMS Tsukuba) using a dry-stamping technique \cite{Castellanos-Gomez2014} onto a 4$\times$4 mm SiO$_2$(90nm)/Si substrate. The surface is covered with tens of hBN flakes in average with varying thickness (few nm up to 100 nm) and size (few $\mu$m up to 200 $\mu$m). Prior to the growth, the substrate is heated in ultrahigh vacuum (UHV) at 800°C during one hour to desorb all the contaminations from the surface, in particular the organic ones left by the mechanical exfoliation process. The base pressure in the molecular beam epitaxy reactor is in the low 10$^{-10}$ mbar range. Molybdenum is evaporated using an electron gun. The deposition rate of Mo controlled by a quartz balance monitor is kept in the 0.0025-0.005 \AA/s range. The Se partial pressure (measured at the sample position thanks to a retractable gauge) is 10$^{-6}$ mbar giving a Se:Mo ratio in the 20-40 range. Considering this large Se:Mo ratio and the volatile character of Se, the deposition rate of Mo mostly determines the deposition rate of MoSe$_2$. The deposited MoSe$_2$ films grown at low temperature ($<$800°C) are systematically annealed at 800°C under Se flux (at 10$^{-6}$ mbar) to improve the crystalline quality. The deposition temperature is varied between 200°C and 800°C. 
After the MBE growth, a second hBN layer (top hBN, a few nm thick)  is transferred on top of the structure by repeating the dry-stamping exfoliation technique. 
Tens of samples of encapsulated MoSe$_2$ monolayers have been fabricated and their properties were investigated by optical spectroscopy. Photoluminescence (PL) and differential reflectivity measurements were performed in a home built micro-spectroscopy set-up around a closed-cycle, low vibration attoDry cryostat with a temperature controller (T = 4–300 K). For PL, a HeNe laser ($\lambda$=633 nm) was used for excitation with a typical power of 30 $\mu$W and integration time of 30 seconds. The white light source for reflectivity measurements is a halogen lamp with a stabilized power supply. The emitted and/or reflected light was dispersed in a spectrometer and detected by a Si-CCD camera. The excitation spot diameter is of the order of 1 $\mu$m. 

%\section{Results and Discussion}

%\begin{figure}[ht!]
%    \centering
%    \includegraphics[width=\linewidth]{Figure1.png}
%    \caption{STEM images for three different MoSe$_2$ coverages on graphene/SiC.}
%    \label{Fig1}
%\end{figure}

\begin{figure}[ht!]
    \centering
    \includegraphics[width=\linewidth]{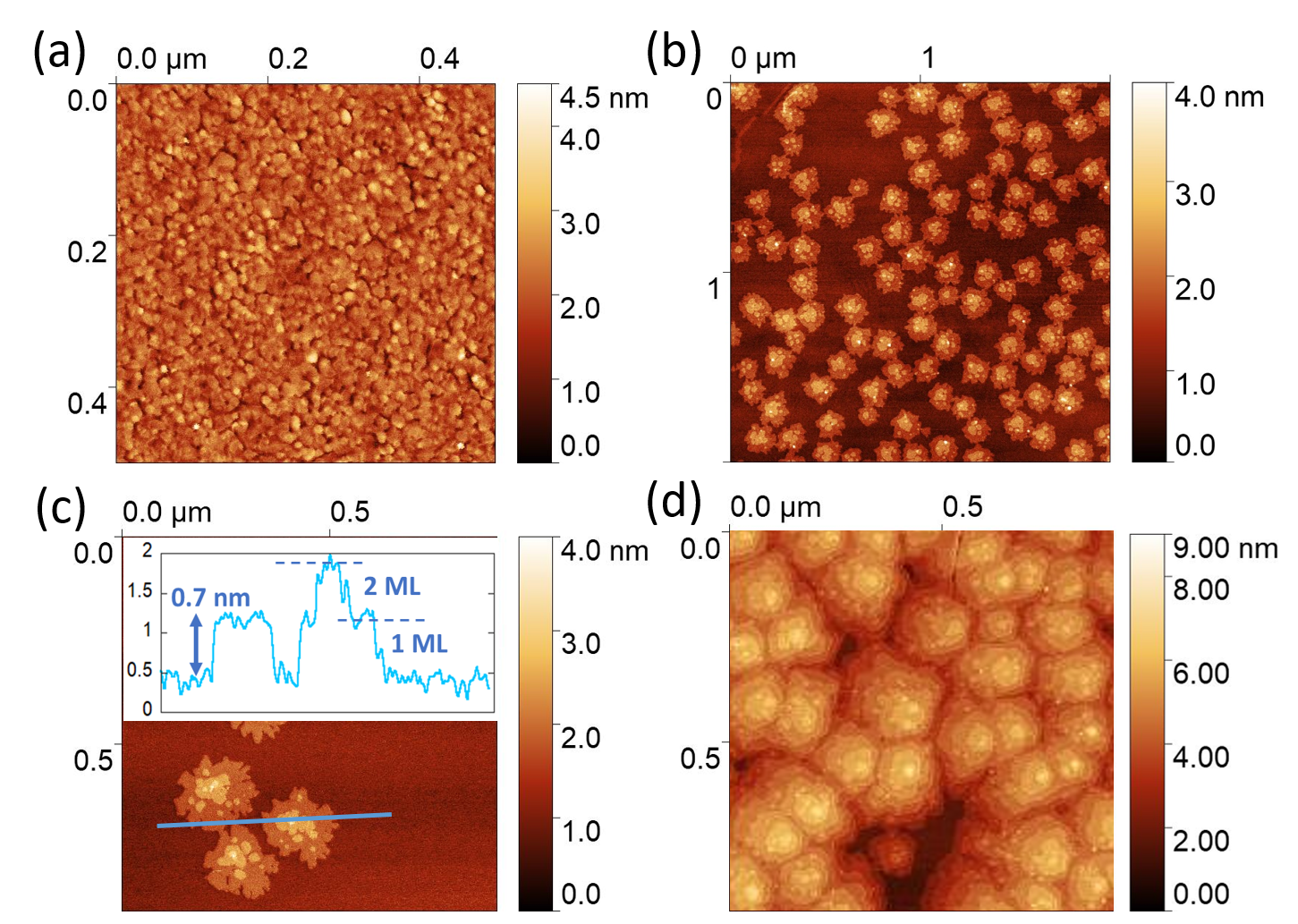}
    \caption{Atomic force microscopy (AFM) image of 1 ML$_{Gr}$ deposited on (a) SiO$_2$ and (b),(c) on hBN. (d) AFM image of 4.5 ML$_{Gr}$ deposited on hBN. The inset in (c) shows the height profile along the blue line, 0.7 nm corresponds to 1 ML of MoSe$_2$ on hBN including the thickness of the MoSe$_2$ triatomic layer and the vdW gap between MoSe$_2$ and hBN.}
    \label{Fig1}
\end{figure}

We first calibrate the MoSe$_2$ deposition rate on graphene/SiC since we thoroughly studied this system in previous works \cite{Khalil2023,Alvarez2018,LeQuang2018,Dau2018}. In this case, the deposition temperature was 300°C followed by in situ annealing during 15 minutes at 800°C under Se flux. The Mo deposition rate is 0.005 \AA/s and the Se partial pressure 10$^{-6}$ mbar. The completion of one monolayer of MoSe$_2$ (1 ML$_{Gr}$) with 100\% coverage corresponds to an equivalent Mo thickness of 2.8 \AA. Using the exact same growth conditions on hBN flakes, we obtain the layer morphology shown in Fig.~\ref{Fig1}a (resp. Fig.~\ref{Fig1}b-c) for 1 ML$_{Gr}$ on SiO$_2$ (resp. hBN).

\begin{figure}[ht!!!]
    \centering
    \includegraphics[width=\linewidth]{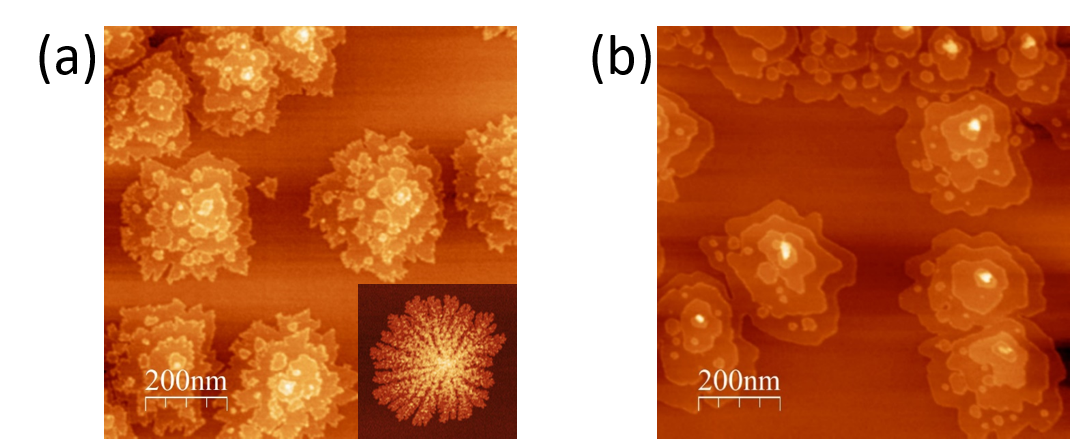}
    \caption{(a) AFM image of 2.25 ML$_{Gr}$ MoSe$_2$ deposited on hBN at 300°C. The inset shows the same deposition but at 200°C. (b) AFM image of the sample in (a) annealed at 800°C during 15 minutes.}
    \label{Fig2}
\end{figure}

On SiO$_2$, we find 100 \% coverage as expected with very small MoSe$_2$ grains with diameters ranging from 10 to 20 nm. However, the coverage on hBN flakes is only $\approx$42 \% with monolayer MoSe$_2$ grains of $\approx$170 nm in diameter. A small fraction of second layer can also be observed in Fig.~\ref{Fig1}c on top of the monolayer grains. Taking into account the first and second layers, the total amount of deposited MoSe$_2$ on hBN flakes as measured by AFM is only $\approx$0.55 ML$_{Gr}$. Hence, the sticking coefficient of MoSe$_2$ monomers (if the reaction between Mo and Se atoms occurs in the gas phase) and/or the residence time of MoSe$_2$ monomers on hBN are clearly less than on graphene and SiO$_2$. Moreover, the low nucleation density demonstrates the much higher monomers mobility on hBN than on graphene or SiO$_2$. In order to reach 100 \% coverage, we deposited 4.5 ML$_{Gr}$ as shown in Fig.~\ref{Fig1}d. In this case, MoSe$_2$ grains coalesce but they are 5 monolayers thick and exhibit a "pyramid" like shape. Based on this result, we then vary the growth temperature T$_g$ with the objective to obtain 100 \% coverage of monolayer MoSe$_2$. First, in order to improve the coverage, we lowered T$_g$ from 300°C down to 200°C keeping the same Mo deposition rate and Se flux to increase the nucleation density. However, as shown in the inset of Fig.~\ref{Fig2}a, as-grown MoSe$_2$ grains exhibit highly dendritic shape which is detrimental to reach 100 \% coverage. Annealing up to 800°C does not change drastically the dendritic shape of the grains. In comparison, as grown grains at 300°C show a compact shape in Fig.~\ref{Fig2}a. We thus consider that T$_g$=300°C is the minimum growth temperature to obtain compact MoSe$_2$ grains. 

\begin{figure}[ht!!!]
    \centering
    \includegraphics[width=\linewidth]{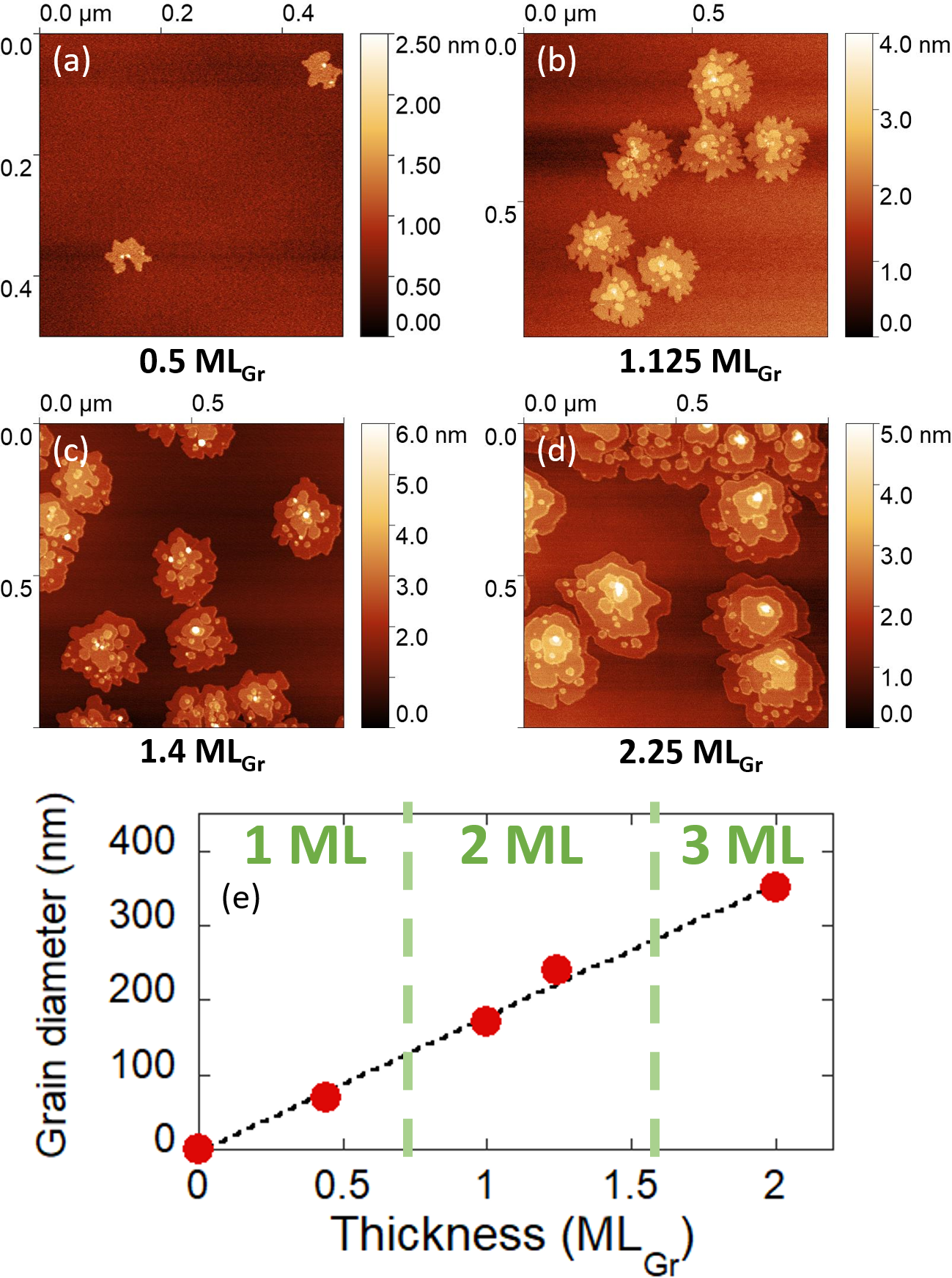}
    \caption{AFM images of (a) 0.5 ML$_{Gr}$; (b) 1.125 ML$_{Gr}$; (c) 1.4 ML$_{Gr}$ and 2.25 ML$_{Gr}$ MoSe$_2$ deposited on hBN at 300°C. (e) Average grain diameter as a function of the deposited thickness in ML$_{Gr}$ extracted from AFM images in (a-d). Based on the observed grain thickness, the graph has been divided into three zones with arbitrary boundaries for 1 ML-thick; 2 ML-thick and 3 ML-thick MoSe$_2$ grains.}
    \label{Fig3}
\end{figure}

The second objective is to stabilize a single layer of MoSe$_2$ on hBN to study optical properties (only monolayer exhibits direct bandgap). Starting from T$_g$=300°C, we first study the effect of annealing to suppress the second and third MoSe$_2$ layers in Fig.~\ref{Fig2}a. The result is shown in Fig.~\ref{Fig2}b: annealing only smoothens the grain edges thanks to higher monomer mobility but the second and third MoSe$_2$ layers remain on top of the grains. The second strategy consists then to increase T$_g$ to avoid the formation of multilayers during the growth. However, when increasing the growth temperature above 400°C, the nucleation density decreases down to zero and hBN flakes are no more covered with MoSe$_2$ after depositing 1 ML$_{Gr}$ (not shown). Finally, we followed a two-step method consisting in depositing 1 ML-thick compact MoSe$_2$ grains at 300°C (nucleation step) and growing the rest of the film up to 1 ML MoSe$_2$ at higher temperature to avoid the formation of multilayers (growth step). For this purpose, we carried out a thorough study of MoSe$_2$ nucleation at T$_g$=300°C varying the content of deposited MoSe$_2$ in ML$_{Gr}$ as shown in Fig.~\ref{Fig3}. It should be noted that all the samples of Fig.~\ref{Fig3} were grown at 300°C and annealed at 800°C during 15 minutes to smoothen the grain edges.

\begin{figure}[ht!!!]
    \centering
    \includegraphics[width=\linewidth]{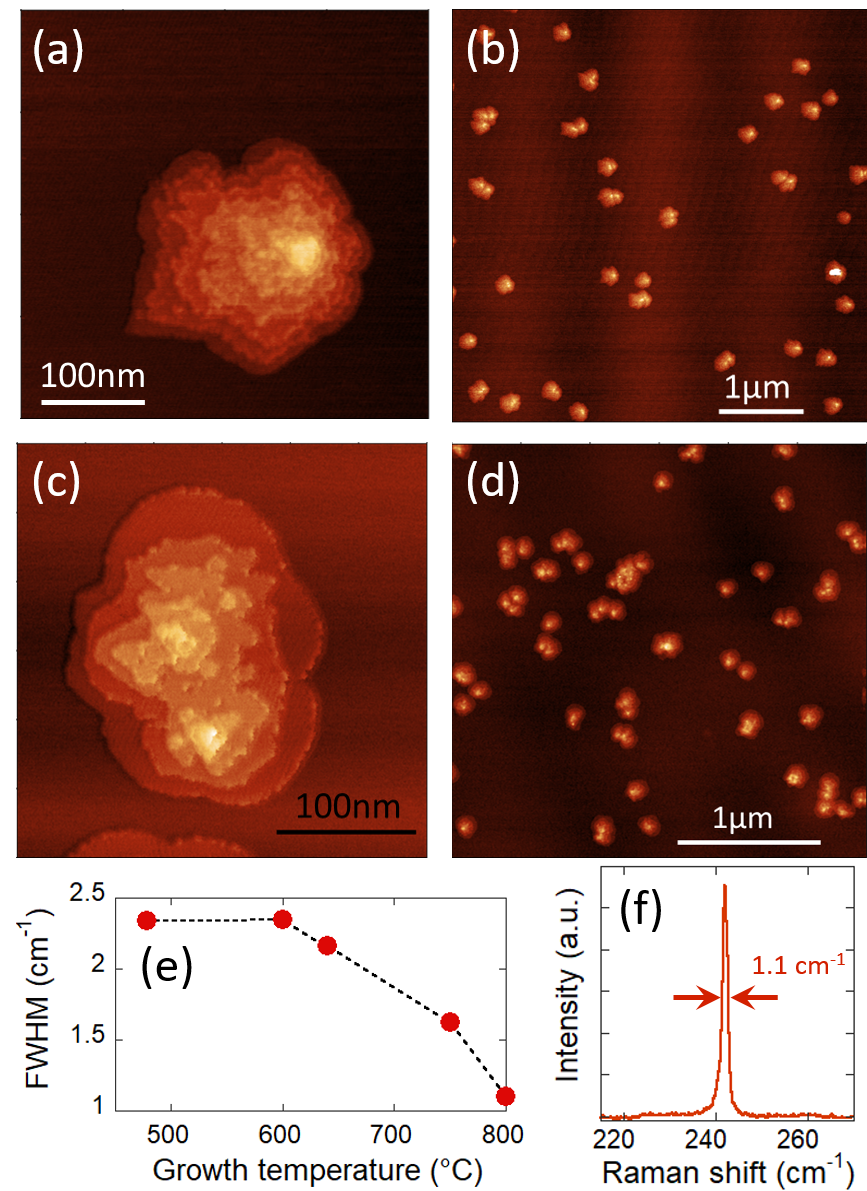}
    \caption{(a), (b) AFM images of 6.1 ML$_{Gr}$ MoSe$_2$ grown at 600°C on hBN. (c), (d) AFM images of 6.5 ML$_{Gr}$ MoSe$_2$ grown at 750°C on hBN. (e) FWHM of the A$_{1g}$ Raman peak of MoSe$_2$ as a function of growth temperature. (f) Raman A$_{1g}$ peak of MoSe$_2$ grown at 800°C.}
    \label{Fig4}
\end{figure}

Increasing the total amount of deposited MoSe$_2$ up to 2 ML$_{Gr}$ in Fig.~\ref{Fig3}a-d clearly increases the thickness of the grains from 1 to 3 ML as plotted in Fig.~\ref{Fig3}e. We conclude that a deposited thickness less than $\approx$0.7 ML$_{Gr}$ is necessary to obtain monolayer thick grains even though their density is very low (2 per 500$\times$500 nm$^2$ in Fig.~\ref{Fig3}a). To ensure monolayer thick MoSe$_2$ grains, we first deposited 0.1 ML$_{Gr}$ of MoSe$_2$ at 300°C and study the effect of growth temperature T$_g$ on the grain thickness and morphology in Fig.~\ref{Fig4}. 

\begin{figure*}[ht!!!]
    \centering
    \includegraphics[width=\linewidth]{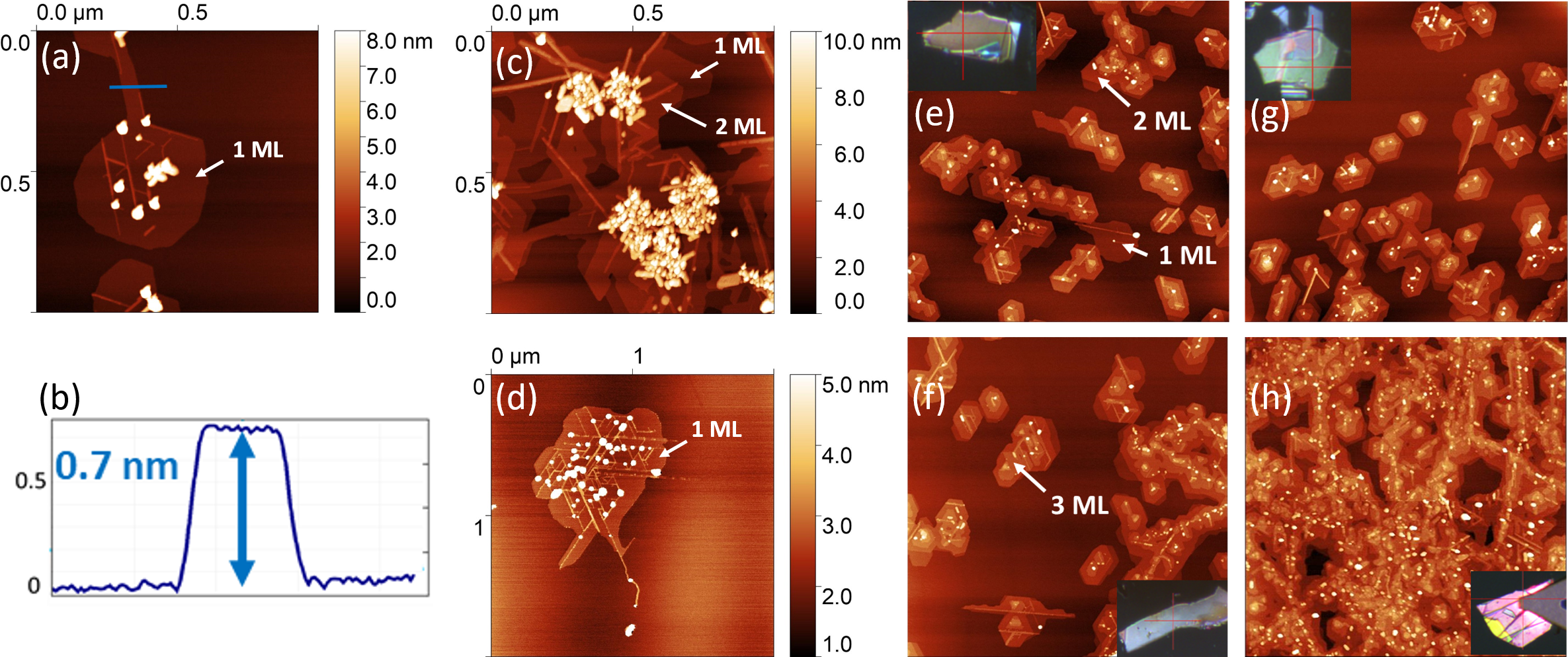}
    \caption{(a) AFM images of 12 ML$_{Gr}$ deposited on hBN at 800°C during 1.5 hours after a nucleation step of 0.1 ML$_{Gr}$ at 300°C (sample 1). (b) Height profile along the blue line in (a) demonstrating the monolayer thickness of the MoSe$_2$ grain. (c) Same MoSe$_2$ deposition but imaged on a second hBN flake. (d) AFM image of 24 ML$_{Gr}$ deposited on hBN at 800°C during 3 hours after a nucleation step of 0.2 ML$_{Gr}$ at 300°C (sample 2). (e-h) AFM images of 3.5 ML$_{Gr}$ deposited on different hBN flakes at 800°C during 55 minutes after a nucleation step of 0.7 ML$_{Gr}$ at 300°C. The insets are optical images of the hBN flakes.}
    \label{Fig5}
\end{figure*}

The results are shown in Fig.~\ref{Fig4}a-b (resp. Fig.~\ref{Fig4}c-d) for T$_g$=600°C (resp. 750°C). Considering the very low sticking coefficient and/or residence time of MoSe$_2$ monomers at the hBN surface at these temperatures, we deposited thicker films of 6.1 ML$_{Gr}$ at 600°C and 6.5 ML$_{Gr}$ at 750°C respectively. For T$_g$=600°C (resp. 750°C), the grains exhibit an average thickness of 5 ML (resp. 4 ML). In order to still reduce the grain thickness down to 1 ML, we selected the highest growth temperature of 800°C. The first remarkable observation is the sharp decrease of the Raman A$_{1g}$ peak full width at half maximum (FWHM) above T$_g$=600°C down to $\approx$1.1 cm$^{-1}$ for T$_g$=800°C as shown in Fig.~\ref{Fig4}e-f. It demonstrates the superior crystalline quality of MoSe$_2$ grains grown at 800°C on hBN flakes. The results for T$_g$=800°C are summarized in Fig.~\ref{Fig5}.

%We varied the nucleation step thickness at T$_g$=300°C from 0.1 ML$_{Gr}$ (Fig.~\ref{Fig6}a-b) to 0.2 ML$_{Gr}$ (Fig.~\ref{Fig6}c) and 0.7 ML$_{Gr}$ (Fig.~\ref{Fig6}d). The growth step was carried out at T$_g$=800°C with a Mo deposition rate of 0.0025 \AA/s and a Se flux of 10$^{-6}$ mbar. On average, after a nucleation step thickness of 0.1 ML$_{Gr}$ at T$_g$=300°C and 1.5 hours of growth corresponding to 12 ML$_{Gr}$, we observe $\approx$500 nm large monolayer thick MoSe$_2$ grains in Fig.~\ref{Fig6}a. They are covered with clusters (bright spots in Fig.~\ref{Fig6}a) and lines of height ... and ... respectively which most probably correspond to Mo-rich nanostructures forming due to the low Se:Mo atomic ratio at the surface of hBN flakes at 800°C. However, we could not decrease (resp. increase) the Mo deposition rate (resp. Se flux) any further with our setup. Fig.~\ref{Fig6}b shows the AFM image of another hBN flake where the coverage is clearly higher with the coexistence of MoSe$_2$ mono- and bilayers along with Mo-rich nanoclusters and nanowires. This illustrates the coverage and thickness dispersion observed from flake to flake. At this stage, we can only speculate that this is due to the hBN flake quality in terms of point defects or steps. After a nucleation step thickness of 0.2 ML$_{Gr}$ at T$_g$=300°C and 3 hours of growth corresponding to 24 ML$_{Gr}$, we observe $\approx$1 $\mu$m large monolayer thick MoSe$_2$ grains in Fig.~\ref{Fig6}c. 

The AFM image of Fig.~\ref{Fig5}a-c (resp. Fig.~\ref{Fig5}d) corresponds to sample 1 (resp. sample 2). In sample 1 (resp. sample 2), MoSe$_2$ was deposited during 1.5 hour (resp. 3 hours) corresponding to a total thickness of 12 ML$_{Gr}$ (resp. 24 ML$_{Gr}$) after 0.1 ML$_{Gr}$ (resp. 0.2 ML$_{Gr}$) nucleation step. The growth step was carried out at T$_g$=800°C with a Mo deposition rate of 0.0025 \AA/s and a Se flux of 10$^{-6}$ mbar. In Fig.~\ref{Fig5}a and ~\ref{Fig5}d, we observe 500 nm and 1 $\mu$m large monolayer thick MoSe$_2$ grains on hBN flakes. They are covered with clusters (bright spots) and lines of height 8 nm and 1 nm respectively which most probably correspond to Mo-rich nanostructures forming due to the low Se:Mo atomic ratio at the surface of hBN flakes at 800°C. However, we could not decrease (resp. increase) the Mo deposition rate (resp. Se flux) any further with our setup. Averaging over larger areas, the coverage is $\approx$5 \% for sample 1 and $\approx$10 \% for sample 2 meaning that the full coverage with 1 ML thick MoSe$_2$ would take approximately 30 hours. In Fig.~\ref{Fig5}c, we show the AFM image of a second flake of sample 1 where the coverage is clearly higher with the coexistence of MoSe$_2$ mono- and bilayers along with Mo-rich nanoclusters and nanowires. This illustrates the coverage and thickness dispersion observed from flake to flake. At this stage, we can only speculate that this is due to the hBN flake surface quality in terms of point defects or steps. Fig.~\ref{Fig5}e-h illustrate the effect of increasing the deposited thickness at the nucleation step to 0.7 ML$_{Gr}$ corresponding to the 1 ML-2 ML thickness transition in the graph of Fig.~\ref{Fig3}e as well as the dispersion of coverage from flake to flake. The MoSe$_2$ grain thickness varies from 1 to 3 ML and the coverage from $\approx$34 \% to almost 100 \%. In summary, ultralow thickness at the nucleation step (0.1-0.2 ML$_{Gr}$) and high temperature growth of 800°C are necessary to stabilize monolayer thick MoSe$_2$ grains on hBN flakes. In the following, we study the optical properties of the samples grown in these conditions.

\begin{figure*}[ht!!!]
    \centering
    \includegraphics[width=\linewidth]{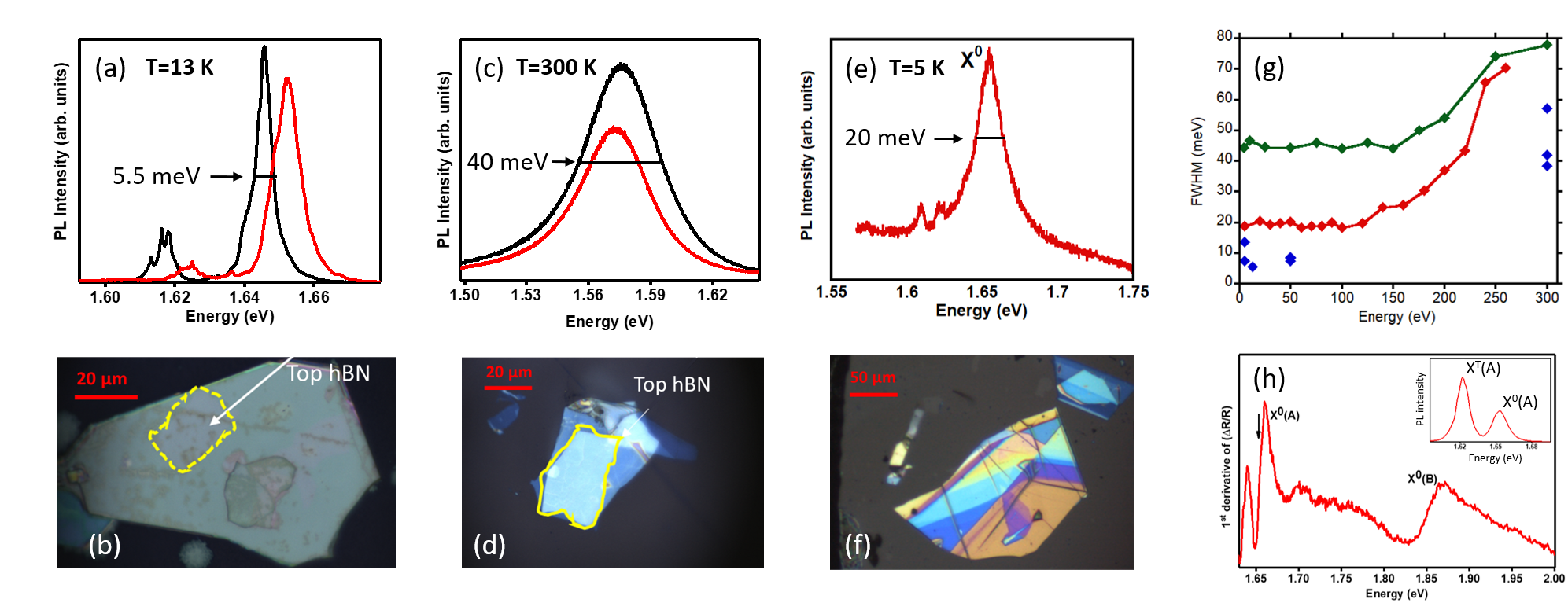}
    \caption{(a) Photoluminescence spectra at T=13 K of hBN/MoSe$_2$/hBN in sample 2 recorded at two different positions in red and black of the hBN flake shown in (b). (c) Photoluminescence spectra at T=300 K of hBN/MoSe$_2$/hBN in sample 1 recorded at two different positions in red and black of the hBN flake shown in (d). (e) PL spectrum recorded at T=5 K on hBN/MoSe$_2$/hBN (shown in (f)) for 1 ML$_{Gr}$ of MoSe$_2$ grown at 300°C and annealed at 800°C. (g) FWHM of PL spectra as a function of temperature for three different samples: sample 2 corresponding to hBN encapsulated MoSe$_2$ grown at 800°C (blue) on two different hBN flakes and different laser spot positions, hBN encapsulated MoSe$_2$ grown at 300°C and annealed at 800°C (red) and MoSe$_2$ directly grown on SiO$_2$/Si \cite{Vergnaud2019} (green). (h) Differential reflectivity of hBN/MoSe$_2$/hBN in sample 1 (shown in (d)) recorded at T=5 K. Inset: corresponding photoluminescence spectrum.}
    \label{Fig6}
\end{figure*}

Fig.~\ref{Fig6}a presents the photoluminescence spectrum for a hBN/MoSe$_2$/hBN structure at T=13 K. It evidences clearly two peaks corresponding respectively to the recombination of neutral exciton (X$^0$) and charged exciton (X$^T$), in agreement with previous results \cite{Cadiz2017,Ajayi2017}. Remarkably the neutral exciton PL shows a narrow linewidth. The Full Width at Half Maximum (FWHM) of the neutral exciton line is $\approx$5.5 meV, demonstrating the high quality of the structure. Although a little larger, this line width is comparable with that obtained with hBN encapsulated MoSe$_2$ monolayer obtained by mechanical exfoliation \cite{Rogers2020,Zhou2020}. The PL spectra do not change much for different points of the sample (see two typical points in Fig.~\ref{Fig6}a). This demonstrates the good growth homogeneity on this hBN flake. Note that the slight change of the PL peak of the neutral and charged exciton observed in Fig.~\ref{Fig6}a also occurs in exfoliated samples (probably induced by small local strain variations). Using the same excitation conditions, ML MoSe$_2$ flakes exfoliated from the bulk crystal exhibit almost two orders of magnitude larger PL intensity. However, as shown in Fig.~\ref{Fig5}, the monolayer coverage is very low in our samples, of the order of few percents, which can partly explain this difference in PL intensity.
In Fig.~\ref{Fig6}c, a reasonable PL intensity at room temperature is obtained and shown at two different positions on the hBN flake. The FWHM is typically 40 meV. For comparison, we show in Fig.~\ref{Fig6}e the PL spectrum at T=5 K of hBN encapsulated MoSe$_2$ grown at 300°C and annealed at 800°C corresponding to the sample of Fig.~\ref{Fig3}b. This spectrum was recorded with a laser energy of 100 $\mu$W and an integration time of 60 seconds. The PL spectrum FWHM for the neutral exciton is $\approx$20 meV demonstrating that growth at high temperature greatly narrows the PL spectrum following the same trend as the FWHM of the A$_{1g}$ Raman peak in Fig.~\ref{Fig4}e. PL spectra FWHM as a function of temperature for three different samples are shown in Fig.~\ref{Fig6}g. For hBN encapsulated MoSe$_2$ grown at 800°C (in blue), the FWHM varies from 5.5 meV at 13 K to 40 meV at room temperature. For hBN encapsulated MoSe$_2$ grown at 300°C and annealed at 800°C (in red), the FWHM varies from 20 meV at 5 K to 70 meV at 250 K. Finally, for non-encapsulated MoSe$_2$ monolayer directly grown on SiO$_2$/Si \cite{Vergnaud2019} (in green), the PL FWHM varies from 45 meV at 5 K to 80 meV at room temperature. We can conclude that high temperature growth on hBN flakes drastically improves the optical properties of MoSe$_2$. Fig.~\ref{Fig6}h displays the differential reflectivity spectrum of a sample grown at high temperature. The peaks corresponding to the absorption of A and B neutral exciton are observed together with the absorption of the charged exciton at lower energy (oscillation below 1.65 eV). The PL peaks of both the neutral and charged exciton are shown in the inset.  This confirms that the MoSe$_2$ monolayer is doped.\\

%\section{Conclusion}
In summary, we studied the growth of MoSe$_2$ on hBN flakes by van der Waals epitaxy in order to obtain monolayer coverage and intense and narrow-linewidth photoluminescence. We found a PL FWHM of 5.5 meV at 13 K and 40 meV at 300 K for the neutral exciton line. These results represent an improvement compared to previous works by MBE. MoSe$_2$ grown by MBE at low temperature (300°C-500°C) shows PL FWHM of 6.6 meV at 10 K (Ref. \citenum{Pacuski2020}) and 60 meV at 300 K (Ref. \citenum{Poh2018}). The linewidth is comparable to the one of MoS$_2$ grown by chemical vapor deposition with a PL FWHM of 5 meV at 4 K (Ref. \citenum{Shree2020}) but still remains a factor two larger than the one of exfoliated MoSe$_2$ flakes encapsulated in hBN ($\approx$2 meV at 4 K in Refs. \citenum{Cadiz2017,Ajayi2017}). To obtain these results, we had to address two main constraints. First, the very low sticking coefficient and/or residence time of MoSe$_2$ monomers at the surface of hBN flakes during the growth leads to very low or even zero nucleation density. To circumvent this issue, we systematically started with a nucleation step at low substrate temperature to obtain monolayer-thick MoSe$_2$ grains. Second, the growth mainly proceeds in a multilayer mode at low and medium substrate temperatures. In order to limit the formation of MoSe$_2$ bilayers, we thus carried out the growth step at very high temperature (800°C). Following this two-step method, we achieved the growth of high quality MoSe$_2$ monolayers with partial coverage exhibiting narrow-linewidth photoluminescence spectra at low temperature and sizeable signal at room temperature. We also demonstrated clear A and B neutral exciton and charged exciton signatures in reflectivity measurements at low temperature. This work demonstrates that molecular beam epitaxy in the van der Waals regime of TMDs on hBN provides high quality crystals comparable to exfoliated ones over large areas. It paves the way for the study of proximity effects in large area and high crystalline quality van der Waals heterostructures.\\

%\begin{acknowledgments}
The authors acknowledge the support from the European Union’s Horizon 2020 research and innovation Programme under grant agreement No 881603 (Graphene Flagship), No 829061 (FET-OPEN NANOPOLY) and No 101079179 (DYNASTY). The French National Research Agency (ANR) is acknowledged for its support through the ANR-18-CE24-0007 MAGICVALLEY and ESR/EQUIPEX+ ANR-21-ESRE-0025 2D-MAG projects. The LANEF framework (No. ANR-10-LABX-0051) is acknowledged for its support through the project 2DMAT. 
%\end{acknowledgments}

\section*{Data Availability Statement}
 The data that support the findings of this study are available from the corresponding author upon reasonable request.

\nocite{*}

\section*{References}

\bibliography{References}% Produces the bibliography via BibTeX.

\end{document}